\documentstyle[12pt,svoss2,psfig]{book}

\begin{document}

\title{Stroboscopic optical observations of the Crab pulsar} 

\author{Simon Vidrih, Andrej \v{C}ade\v{z}, Mirjam Gali\v{c}i\v{c}}

\affil{Department of Physics, Faculty of Mathematics and Physics, University of
Ljubljana, Jadranska 19, 1000 Ljubljana, Slovenia}

\author{Alberto Carrami\~nana}

\affil{I.N.A.O.E., Luis Enrique Erro 1, Tonantzintla, Puebla 72840, M\'exico}

\begin{abstract}
Photometric data of the Crab pulsar, obtained in stroboscopic mode over a
period of more than eight years, are presented here. The applied Fourier
analysis reveals a faint 60 second modulation of the pulsar's signal, which we
interpret as a free precession of the pulsar.
\end{abstract}


\vspace{1cm}

\section{Introduction}

Among all known pulsars, the Crab pulsar is the brightest in all the spectral
regions and, therefore, by far the most thoroughly investigated. It is $\sim
1000$ year old, with the rotation period of $\sim 33\,$ms. In optical it is a
16.5 magnitude star and its pulsations were detected more than 30 years ago
(Cocke et al. 1969). We observed the Crab pulsar optically in stroboscopic mode
for more than eight years. We detected a faint peak in Fourier spectra of the
pulsar's light curve at frequency $\sim 1/60\,$Hz and suggested that it may be
the signature of free precession of the pulsar. Another statistical analysis
applied on the extensive data set confirms the presence of a periodic signal
with the period of 60 seconds.

\section{Stroboscopic observations of the Crab pulsar}

Our stroboscopic system is based on a shutter that
opens with the prescribed frequency and phase. The light signal is thus observed
periodically only during one part of the period. We have applied this technique
in performing highly accurate phase resolved optical photometry of the Crab
pulsar.

\begin{figure}[t]
\centerline{\hbox{
\psfig{figure=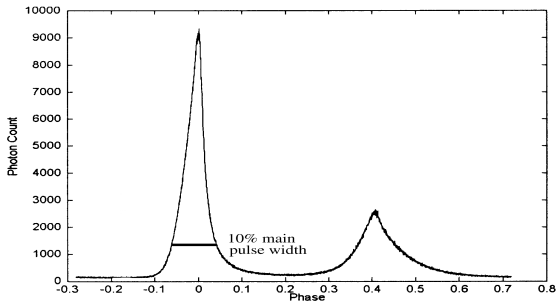,height=4.5 cm,width=8 cm}
}}
\caption{The phase light curve of the Crab pulsar as taken with the fast speed
photometer on board of the Hubble Space Telescope (Percival et al. 1993).
\label{vidrih_fig1.eps}}
\end{figure}

The phase light curve of the Crab pulsar is well known (see Figure
\ref{vidrih_fig1.eps}) and the arrival times of the pulses can be precisely
calculated. The shutter is a rotating wheel with the width of
the out-cuts corresponding to the width of the Crab pulsar's main pulse. While
observing, we lock our stroboscopic shutter in phase to the main peak, so
that the CCD detector receives all light from the main pulse, but the bright
emission from the surrounding nebula is reduced by a factor of 10, which
substantially lowers the noise in the pulsar's signal. In the stroboscopic mode
we take several hundred consecutive images with exposure times of some
seconds and thus obtain, at each observing run, the pulsar's light curve
$m_k(t_l)$, where $t_l$ are consecutive starting times of exposures in the k-th
observing run.

A faint 60 second periodic modulation of the Crab pulsar light curve was
identified already seven years ago (\v{C}ade\v{z} \& 
Gali\v{c}i\v{c} 1996) and a simple free precession model was proposed to explain
the observed phenomenon (\v{C}ade\v{z} et al. 1997). We have continued
to monitor the pulsar and up to the year 2001 a set of 3400 images has been
built. It covers more than 20 hours of photometry data (exposure times of the
images are between 4 and 15 seconds and sampling rates are between $4.0$ and
$26.7$ seconds) taken with four different telescopes (the Hubble Space
Telescope, the $1.22\,$m and $1.82\,$m telescopes of the Asiago Observatory and
the $2.12\,$ telescope of the Guillermo Haro Observatory). Data are spanning a
period of almost nine years (\v{C}ade\v{z} et al. 2001).

\section{Data analysis and results}

The following Fourier analysis on our data is applied to confirm for the
observed 60 second modulation. For each obtained
light curve $m_k(t_l)$ we calculate its Fourier transform
\begin{equation}
F_k(\omega)={1\over \sqrt{2 \pi (t_{l_{max}}-t_{l_{min}})}} \sum_l
m_k(t_l)~e^{i\omega t_l}~ {{t_{l+1}-t_{l-1}}\over 2}
\end{equation}
in order to test if all Fourier transforms contain a common signal at a given
frequency. However, since the rotation frequency $\omega_{rot}$ of the pulsar is
slowly decreasing one should expect that the free precession frequency
$\omega_{fp}$ should also be decreasing. Therefore we have to allow for the
recalibration of frequency scale to a common date in order to take this effect
into account. The theoretical prediction (\v{C}ade\v{z} et al. 1997) is that
$\omega_{fp}$ can change as $\omega_{rot}^p$, where p can go from 1 to 3. Thus
we rescale all the Fourier transforms to the same date (Jan. 1, 1996) as
\begin{equation}
\tilde F_k(\omega)=F_k \left(\omega
\left(\frac{\omega_{rot}^{\circ}}{\omega_{rot}}\right)^p\right)\ ,
\end{equation}
where $\omega_{rot}^{\circ}$ is the pulsar's rotation frequency at the reference
date and $\omega_{rot}$ is the pulsar's rotation frequency at the time of
observations. For the value of $p$ we take 1 (\v{C}ade\v{z} et al. 2001).

With $\tilde F_k (\omega)$ we construct a matrix that has for its components
cross-correlation functions between pairs of Fourier transforms
\begin{equation}
M_{jk}(\omega)	=\tilde F_j(\omega)\tilde F_k(\omega)\ .
\end{equation}
If $m_k$ were white statistically independent random processes, then
$M_{jk}(\omega)$ for all $j$, $k$ and given $\omega$ would be a random process
distributed in the complex plane according to a $2D$ Gaussian distribution. If,
however, $M_{jk}(\omega)$ contain a common signal , then they are distributed in
a ring around the origin of the complex plane. The probability that a common
signal is present is large if the effective width of the ring is significantly
smaller than its radius.

\begin{figure}[ht]
\centerline{\hbox{
\psfig{figure=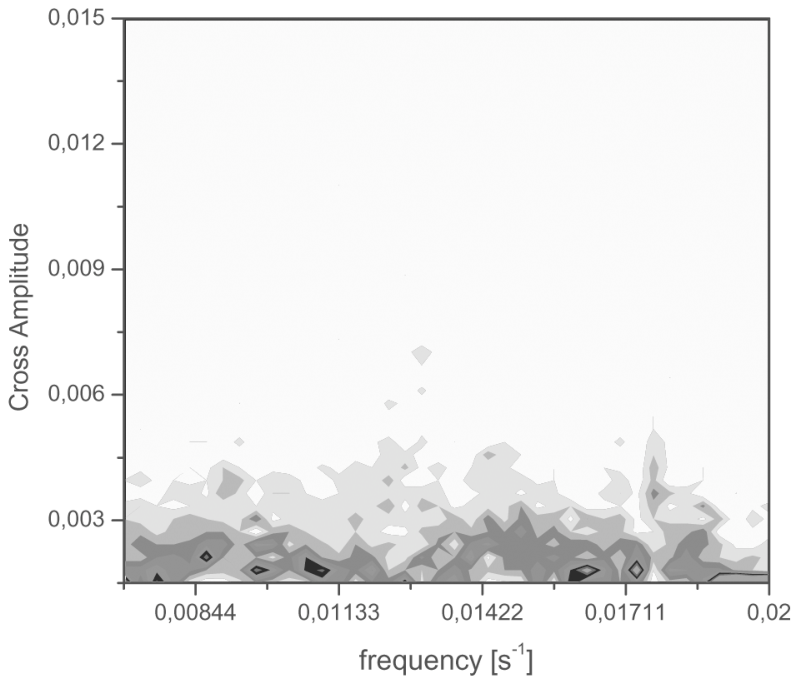,height=5 cm,width=6.6 cm} 
\psfig{figure=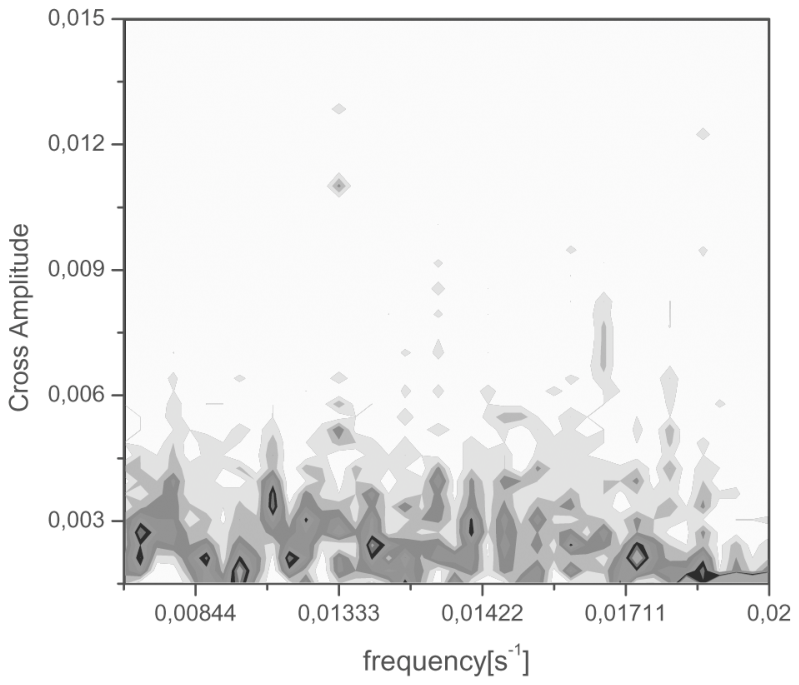,height=5 cm,width=6.6 cm}
}}
\caption{Contour plot representing radial distributions of cross-correlation
functions as a function of frequency for the test star (left) and the Crab
pulsar (right). The x-axis represents the recalibrated frequency in Hz and
y-axis represents the amplitude of the cross-correlation function in
$\mathrm{mag^2/Hz}$. Number of $M_{jk}$ binned in 45 bins is presented with
different shades of gray, the brighter is the color the lower is the number.
\label{vidrih_fig23.eps}}
\end{figure}

At a discrete sample of frequencies $\omega_s$, where
$\nu_s=\omega_s/2\pi$ goes from $0.007$ Hz to $0.02$ Hz in steps of $3.3\times
10^{-4}$Hz, we calculate radial distributions $\frac{dN}{d|M_{jk}|}(\omega_s)$
of cross-correlation functions $M_{jk}(\omega_s)$. The number of
cross-correlated pairs binned in 45 bins of width $3.3\times
10^{-4}\mathrm{mag^2/Hz}$ is 136 in the case of the Crab pulsar and 78 in the
case of a test star. The results for the test star and the Crab pulsar are shown
in Figure \ref{vidrih_fig23.eps}. In the case of the test star all distributions
are similar which confirms the assumption of white Gaussian noise.
It is apparent that the Crab pulsar is noisier than the test star which is also
in agreement with \v{C}ade\v{z} et al. (2001). At a frequency 
$\sim 0.0167$ Hz we can see brighter colors on the contour plot for the values
of $|M_{jk}|$ between 0 and 0.004 $\mathrm{mag^2/Hz}$ on one side and `a darker
island' for the values of $|M_{jk}|$ around 0.007 $\mathrm{mag^2/Hz}$. It is 
clear, that a periodic signal at the frequency $\sim 1/60\,$Hz is
present.

\section{Conclusions}

Presented evidence suggests that the Crab pulsar free precesses. However, the
measured free precession signal is too faint to give a final answer whether
the amplitude of the free precession changes with time and on what time scale
the pulsar relaxes its internal stress. Examining the free precession one can
learn about the equation of state of the neutron matter in pulsars. Moreover,
Chandra observations of the Crab pulsar and its nebula (Weisskopf et al. 2000)
has given additional motivation to understand interaction mechanisms between the
pulsar and highly magnetized plasma around it. We are continuing to
observe the Crab pulsar in stroboscopic mode. The increase in
signal to noise ratio of the free precession signature will help answer some
of the open questions regarding pulsar physics.



\begin{references}

\reference Cocke, W. J., Disney, M. J., and Taylor, D. J. 1969, Nature, Lond.
221, 525.

\reference \v{C}ade\v{z}, A., and Gali\v{c}i\v{c}, M. 1996, \aap, 306, 443.

\reference \v{C}ade\v{z}, A., Gali\v{c}i\v{c}, M., and Calvani, M. 1997, \aap,
324, 1005.

\reference \v{C}ade\v{z}, A., Vidrih, S., Gali\v{c}i\v{c} M., and Carrami\~nana,
A. 2001, \aap, 366, 930.

\reference Percival, J. W., et al., 1993, \apj, 407, 276.

\reference Weisskopf, M. C., et al., 2000, \apj, 536L, 81.

\end{references}
\end{document}